\documentclass[aps,prl,amsmath,twocolumn,superscriptaddress,showpacs,floatfix]{revtex4}

\def\CeAl{CeAl$_3$} 
\def\CeYAl{Ce$_{1-x}$Y$_x$Al$_{3}$} 
\def\CeLaAl{Ce$_{1-y}$La$_y$Al$_{3}$} 
\def\Chi{$\mathrm{Im}\,\chi(\omega)$}

\def\Vkf{V$_{kf}$}

\newif\ifpdf
\ifx\pdfoutput\undefined
\pdffalse 
\else
\pdfoutput=1 
\pdftrue
\fi

\ifpdf
\usepackage[pdftex]{graphicx}
\else
\usepackage{graphicx}
\fi

\begin{document}

\title{The Transition from Heavy Fermion to Mixed Valence in \CeYAl:\\ 
A Quantitative Comparison with the Anderson Impurity Model}
\date{\today}
\author{E. A. Goremychkin}
\affiliation{Materials Science Division, Argonne National Laboratory, Argonne, IL 60439}
\affiliation{ISIS Pulsed Neutron and Muon Source, Rutherford Appleton Laboratory, OX11 0QX, UK}
\author{R. Osborn}
\email{ROsborn@anl.gov}
\affiliation{Materials Science Division, Argonne National Laboratory, Argonne, IL 60439}
\author{I. L. Sashin}
\affiliation{Joint Institute for Nuclear Research, Dubna, Moscow Region, 141980 Russia}
\author{P. Riseborough}
\affiliation{Department of Physics, Temple University, Philadelphia, PA 19122, USA}
\author{B. D. Rainford}
\affiliation{Department of Physics, University of Southampton, S017 1BJ, UK}
\author{D. T. Adroja}
\affiliation{ISIS Pulsed Neutron and Muon Source, Rutherford Appleton Laboratory, OX11 0QX, UK}
\author{J. M. Lawrence}
\affiliation{Department of Physics, University of California, Irvine, CA 92697, USA}

\ifpdf
\DeclareGraphicsExtensions{.pdf, .jpg, .tif}
\else
\DeclareGraphicsExtensions{.eps, .jpg}
\fi

\begin{abstract}
We present a neutron scattering investigation of \CeYAl\ as a function of chemical pressure, which induces a transition from heavy fermion behavior in \CeAl\ (T$_\mathrm{K}=5$ K) to a mixed valence state at $x=0.5$ (T$_\mathrm{K}=150$ K).  The crossover can be modeled accurately on an absolute intensity scale by an increase in the \textit{k-f} hybridization, V$_{kf}$, within the Anderson Impurity Model. Surprisingly, the principal effect of the increasing \Vkf\ is not to broaden the low-energy components of the dynamic magnetic susceptibility but to transfer spectral weight to high energy.
\end{abstract}


\maketitle

The Anderson Impurity Model (AIM) has been invoked to describe the thermodynamic, transport, and spectroscopic properties of strongly correlated electron systems for nearly fifty years \cite{Anderson:1961p24172}. It is believed to contain the essential physics of materials in which a narrow band of quasi-localized electrons are hybridized with a broader band of itinerant electrons, and provides a mechanism describing the formation and screening of local moments in metallic systems. The AIM has been most widely applied to describe fluctuating moments in rare earth $f$-electron systems, ranging from strongly hybridized mixed valent systems to more weakly hybridized heavy fermion systems \cite{Fulde:1988p24212}. Although it is a single-impurity model, it has been successfully applied to concentrated rare earth systems, except at the lowest temperatures where lattice coherence cannot be ignored. 

In spite of its widespread use in heavy fermion and mixed valence physics, there have been no direct quantitative comparisons of \textit{ab initio} calculations of the dynamic magnetic susceptibility, \Chi, with experiment covering the crossover between the two regimes. In recent years, it has become possible to perform detailed calculations of \Chi\ using the Non-Crossing Approximation (NCA) in the presence of crystal field splittings of the 4$f$-electron ground multiplet \cite{Bickers:1987p24211,Kuramoto:1983p24126}, but these theoretical advances have never been tested against inelastic neutron scattering data, which directly measure \Chi. The purpose of this letter is to present the first detailed quantitative comparison of AIM/NCA calculations with neutron scattering data as a function of the hybridization strength, \Vkf, covering the transition from heavy fermion to mixed valence behavior. There is excellent agreement on an absolute intensity scale between theory and experiment, although the relative strength of the transverse and longitudinal susceptibility requires an adjustment for anisotropic hybridization. The low-frequency broadening of the magnetic response is typically proportional to the Kondo temperature, T$_\mathrm{K}$, so it is surprising that increasing the hybridization does not significantly increase the width of the low-energy contributions to \Chi. Instead, the transition is predominantly characterized by a transfer of spectral weight from the two low-energy components, respectively a narrow quasielastic response and a crystal field excitation, to a broad high-energy tail.

\begin{figure}[b]
\begin{center}
\vspace{-.2in}
\centerline {
\includegraphics[width=2.0in]{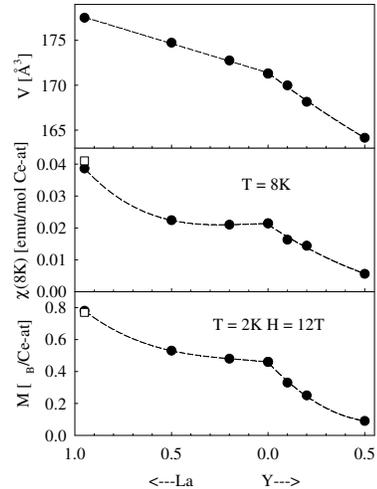}
}
\caption{The unit cell volume, the static susceptibility measured at 8K and the magnetization in a field of 12T measured at 2K of  \CeLaAl\ and \CeYAl. The solid circles are measurements and the open squares are predictions of a localized crystal field model. The lines are guides to the eye. }
\vspace{-.25in}
\end{center}
\end{figure}

\begin{figure*}[t]
\begin{center}
\vspace{-.3in}
\centerline {
\includegraphics[width=5.5in]{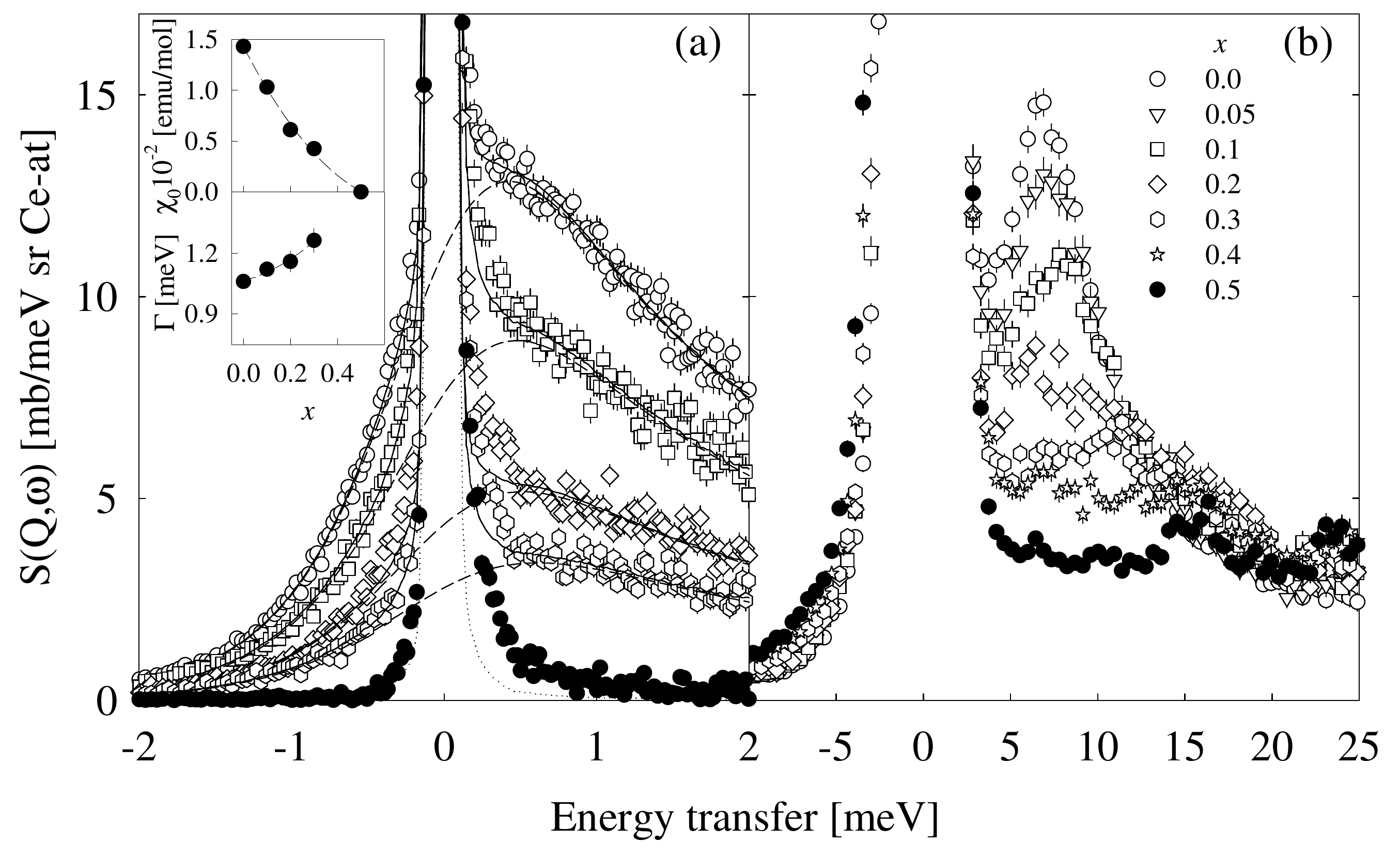}
}
\caption{S(Q,$\omega$) of \CeYAl\ as a function of $x$ measured at an average scattering angle of (a) $60^\circ$ and (b) $15^\circ$, with an incident energy of (a) 3.12 meV and (b) 35 meV. The solid lines are fits to a quasielastic Lorentzian lineshape (dashed line) and an elastic peak (dotted line) convolved with the instrumental resolution (FWHM $\sim 90\mu$eV). The inset shows the fitted values of the static susceptibility, $\chi_0$, and the half-width of the quasielastic peak, $\Gamma$.}
\vspace{-.4in}
\end{center}
\end{figure*}

The variation in hybridization is achieved through the application of chemical pressure on the canonical heavy fermion system \CeAl \cite{Andres:1975p24245}. Doping lanthanum onto the cerium sites expands the lattice, progressively localizing the 4$f$ electrons while doping with yttrium compresses the lattice, inducing a transition from the heavy fermion behavior of pure \CeAl, characterized by a Kondo temperature of T$_\mathrm{K} = 5$K, to a mixed valence state in Ce$_{0.5}$Y$_{0.5}$Al$_3$, where T$_\mathrm{K}= 150$K.  Comparison with the cell volumes of pure \CeAl\ under pressure shows that the chemical pressure in \CeYAl\ at $x = 0.5$ is equivalent to 27 kbar. The bulk measurements in Figure 1 illustrate how chemical pressure, and the consequent increase in hybridization, reduces the static susceptibility and high-field magnetization from the values predicted in a pure crystal field model  \cite{Goremychkin:1999p63}. 

Polycrystalline samples of \CeYAl, with $x$ = 0.0, 0.05, 0.1, 0.2, 0.3, 0.4, and 0.5, \CeLaAl\ with $y$ = 0.2, 0.3, 0.4, 0.5, and 0.95, along with non-magnetic samples of LaAl$_3$ and La$_{0.5}$Y$_{0.5}$Al$_3$, were made by arc-melting stoichiometric quantities of the constituent elements, with a total mass of approximately 30 g for each sample, followed by annealing in a vacuum at 900$^\circ$C for four to five weeks. Powder neutron diffraction confirmed that each sample was single-phase. The inelastic neutron scattering experiments were performed on the IN6 spectrometer at the Institut Laue Langevin, using an incident neutron energy of 3.12 meV, and the LRMECS spectrometer at the Intense Pulsed Neutron Source, in Argonne National Laboratory, using incident energies of 35 and 80 meV. All spectra were normalized on an absolute intensity scale using a standard vanadium plate and corrected for self-shielding, absorption, and form factor. Our own comparison with inelastic scattering standards has shown that it is possible to achieve accuracies of 1\% or better, a precision that is confirmed here in our comparisons of data taken at different incident neutron energies.

Figure 2 shows corrected data from IN6 and LRMECS at a temperature of 8 K covering the low and intermediate energy range. In Figure 2(a), the magnetic response is fitted to a single quasielastic Lorentzian lineshape convolved with the instrumental resolution. 
\begin{eqnarray}
 \label{SQw}
  S(Q,\omega) \propto
  F^2(Q) \left[n(\omega)+1\right]  \frac{\chi_0}{2\pi}  \frac{\omega\Gamma}{\left(\omega^2 + \Gamma^2 \right)}
\end{eqnarray}
where $F(Q)$ is the Ce$^{3+}$ magnetic form factor, $n(\omega)$ is the Bose population factor, $\chi_0$ is the static susceptibility, and $\Gamma$ the half width. 

With increasing $x$, there is only a modest increase in the linewidth $\Gamma$ from 1.06 meV at $x = 0$ to 1.26 meV at $x = 0.3$, but a dramatic reduction in the spectral weight, $\chi_0$, becoming negligibly small at $x = 0.5$. The energy range of Fig. 2(b) covers the crystal field excitation, which is observed at 6.7 meV in \CeAl. With increasing hybridization, the crystal field peak falls sharply in intensity, with a slight shift to higher energy, and is difficult to resolve at $x\geq0.3$.

\begin{figure}[b]
\vspace{-.3in}
\begin{center}
\centerline {
\includegraphics[width=3in]{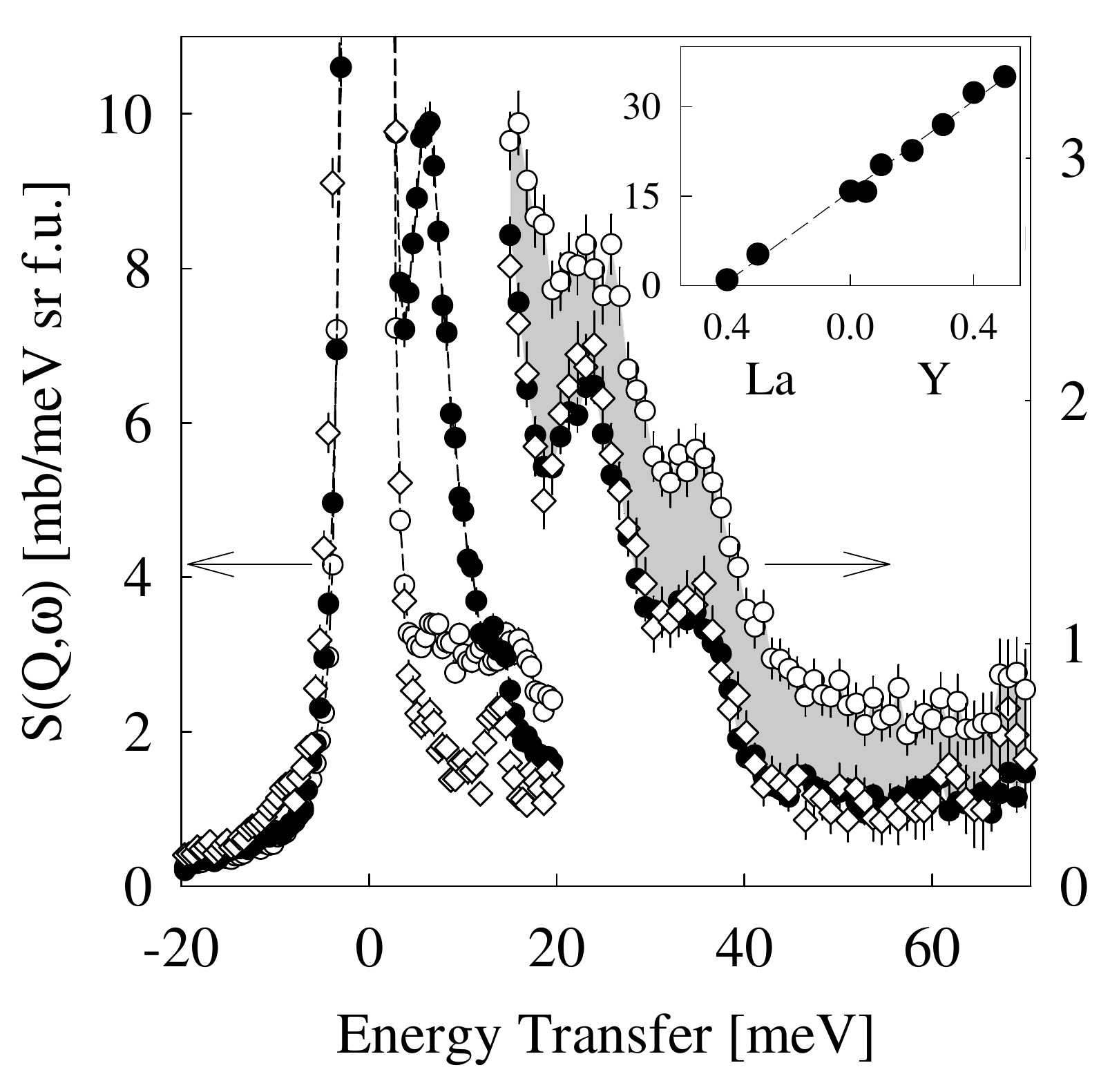}
}
\caption{S(Q,$\omega$) of Ce$_{0.6}$Y$_{0.4}$Al$_3$ (open circles), Ce$_{0.6}$La$_{0.4}$Al$_3$ (solid circles), and LaAl$_3$ (diamonds) measured at an average scattering angle of $15^\circ$ with an incident energy of 35 meV (left axis) and 80 meV (right axis) at 8 K, normalized on an absolute scale.  The shaded area shows the estimated magnetic scattering. The inset shows the magnetic scattering integrated from 40 to 70 meV as a function of La and Y doping.}
\vspace{-.5in}
\end{center}
\end{figure}

At all values of $x$, there is an additional component to the magnetic response in the form of a high-energy tail, extending to greater than 70 meV, that cannot be accounted for by the low-energy quasielastic Lorentzian or the crystal field excitation. It is also just visible at $y = 0.3$, but it has negligible intensity at $y = 0.4$. At these energies, multiple scattering produces a significant phonon contribution to the measured inelastic scattering and must be subtracted before comparing to theory. Figure 3 illustrates this with data at $x = 0.4$, which shows significant additional magnetic intensity when compared to the non-magnetic LaAl$_3$ data. The inset to Figure 3 shows that the high-energy magnetic component, integrated from 40 to 70 meV, falls with increasing cell volume, \textit{i.e.}, with decreasing hybridization, becoming negligible in Ce$_{0.6}$La$_{0.4}$Al$_3$. 

\begin{figure}[htb]
\begin{center}
\centerline {
\includegraphics[width=3in]{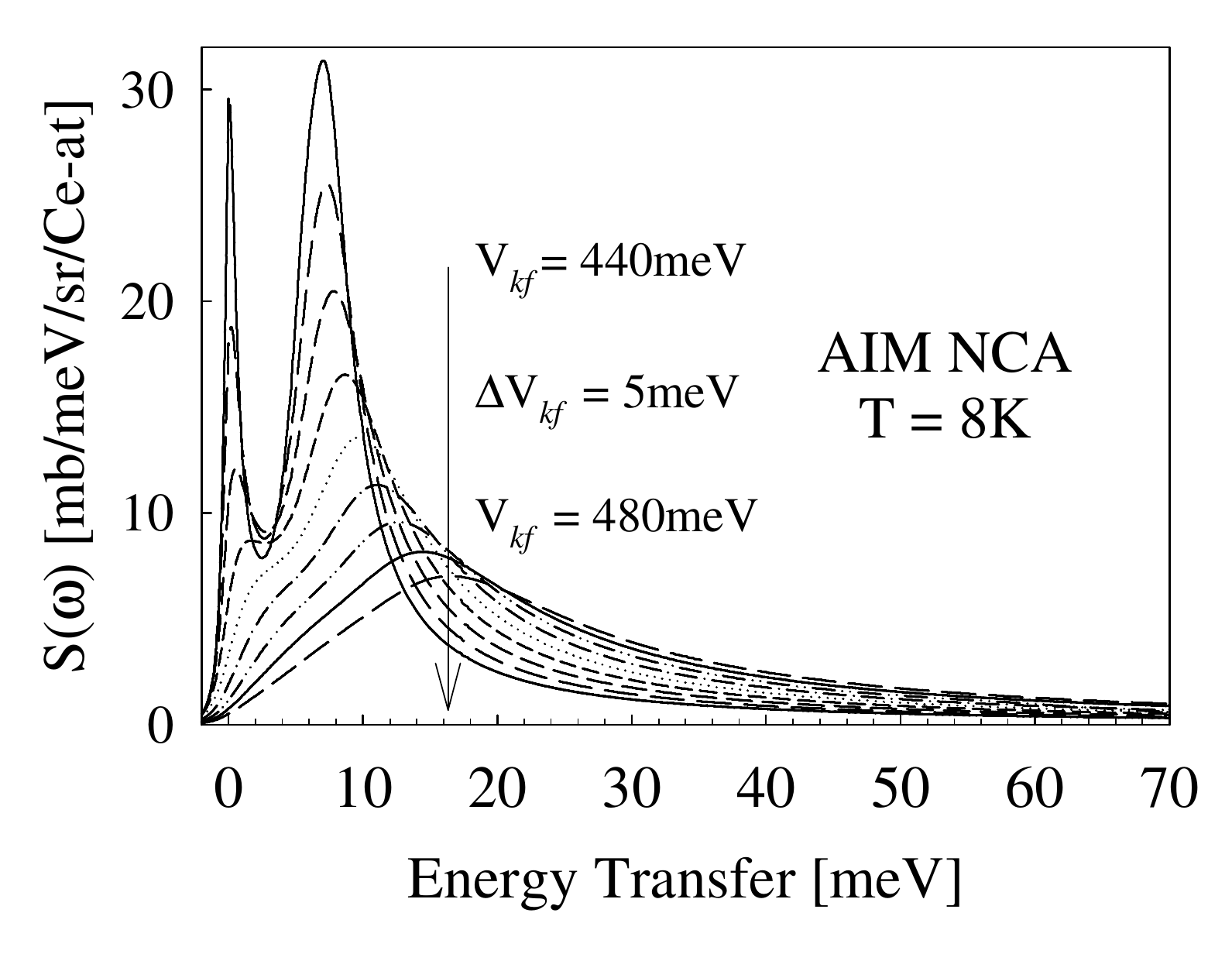}
}
\caption{Theoretical calculations of S($\omega$) determined by the Anderson Impurity Model using the Non-Crossing Approximation at a temperature of 8 K.}
\vspace{-.45in}
\end{center}
\end{figure}

Qualitatively, the observations are consistent with AIM/NCA calculations, which follow the method of Kuramoto \cite{Kuramoto:1983p24126} and Bickers {\it et al.} \cite{Bickers:1987p24211} (Figure 4). We assume a Gaussian conduction band of half-width $W=3$ eV, as in Ref. \cite{Riseborough:2003p3703} with the ground state of the Ce$^{3+}$ 4$f$ electrons at $E_f=-2$ eV. The 14-fold degeneracy of the $f$-states is lifted by the free-atom spin-orbit coupling and the crystal field potential, which, in hexagonal point group symmetry, is diagonal in $J_z$. The $J = \frac{5}{2}$ multiplet is therefore split into three Kramers doublets, $\Gamma_7 = \left|\pm\frac{1}{2}\right>$, $\Gamma_8 = \left|\pm\frac{5}{2}\right>$, and $\Gamma_9 = \left|\pm\frac{3}{2}\right>$.  In an earlier publication, we showed that the $\Gamma_9$ doublet is the ground state, with the other two doublets nearly degenerate at an energy of approximately 7 meV \cite{Goremychkin:1999p63}; in the calculation, the $\Gamma_8$ and $\Gamma_7$ doublets were assumed to be at 6 meV and 7.5 meV, respectively.  The hybridization matrix element was assumed to be isotropic and its value was treated as a free parameter in order to simulate the effect of increasing $x$.  

For low values of the hybridization, the AIM/NCA calculations show a sharp quasielastic response, well-defined crystal field excitation, and a broad high energy tail. The high-energy tail in the spectrum is a consequence of the dynamic screening of the local moments by the conduction electrons \cite{Anderson:1967p24164}. The screening of the magnetic moments can be considered as an iterated orthogonality catastrophe \cite{Anderson:1967p24165}, which occurs since the conduction electrons experience a sequence of local spin-dependent potentials caused by successive flips of the local spin. This process results in a power-law decay of the spectrum at high energies, similar to that found in the x-ray edge problem \cite{Mahan:1967p24166}.

As the hybridization increases, the calculated quasielastic and crystal field contributions gradually decrease in intensity, but their spectral weight is transferred into the high energy tail. This agrees with the experimental trends with increasing $x$. Since the theoretical predictions cannot be described by simple analytic functions, we have put this comparison on a quantitative footing by directly comparing integrals over energy ranges corresponding to the three components in the scattering with both the experimental and calculated intensities placed on an absolute scale with no adjustable parameters. The quasielastic component is represented by an integral from -3 to 2 meV (using the fitted profile to exclude the elastic nuclear scattering), the crystal field component from 5 to 15 meV, and the high energy component from 40 to 70 meV. For the latter two components, similar integrals were performed on LaAl$_3$ and La$_{0.5}$Y$_{0.5}$Al$_3$, and subtracted after interpolating for the correct yttrium concentration. The integrals in the high-energy tail are not affected by this interpolation since the phonon scattering at these energies is dominated by the aluminum contribution.

\begin{figure}[t]
\begin{center}
\centerline {
\includegraphics[width=2.5in]{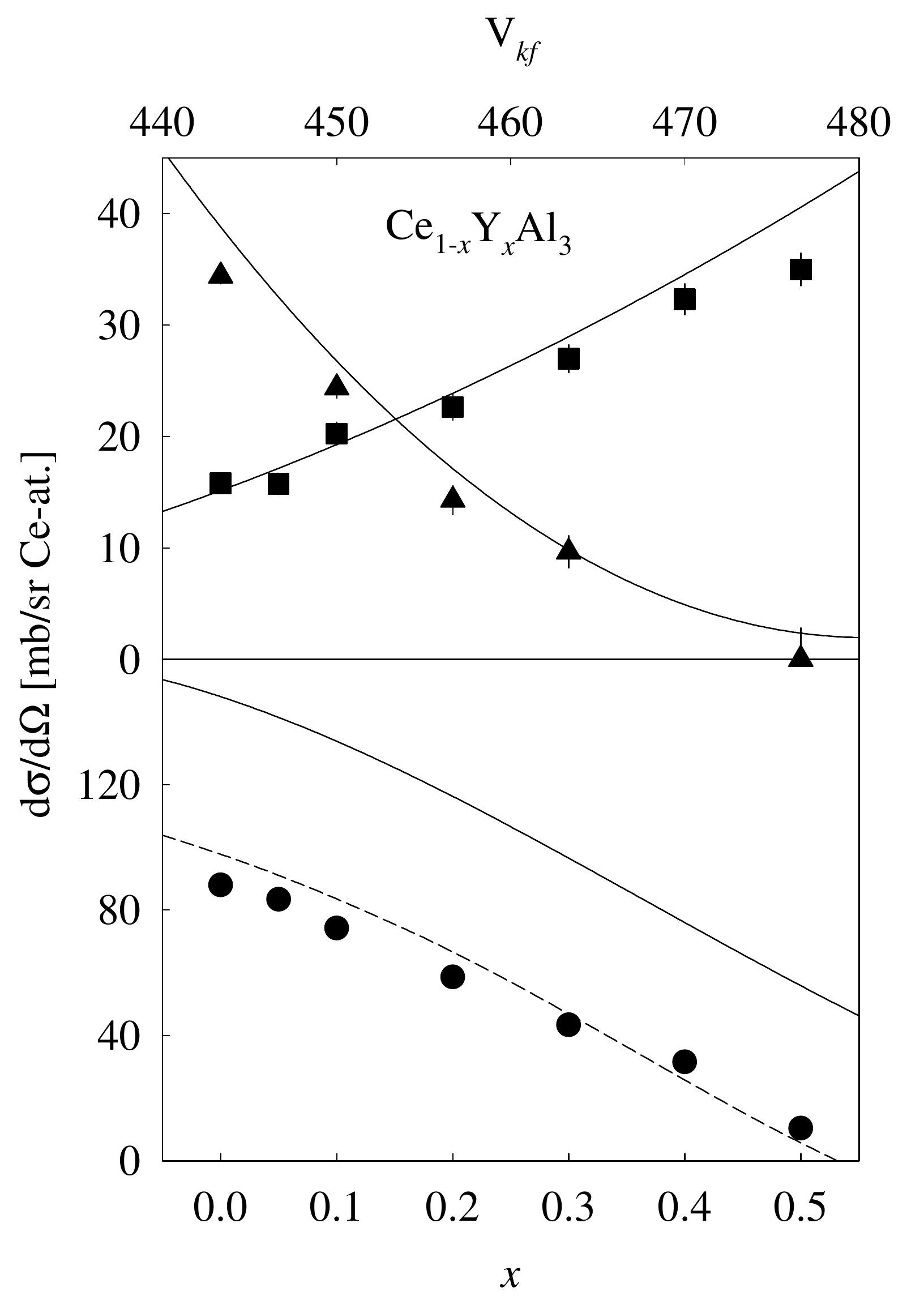}
}
\caption{The magnetic cross section of \CeYAl\ as a function of $x$ integrated over three energy regions, -3 to 2 meV (triangles), 5 to 15 meV (circles) and 40 to 70 meV (squares), and corrected for the form factor. The solid lines are the same integrals determined from the theoretical calculations using the AIM/NCA model as a function of \Vkf, using a linear mapping of \Vkf\ to $x$. The dashed line is the same as the solid line shifted to illustrate the effect of increasing \Vkf\ by $\sim20$ meV at each value of $x$.}
\end{center}
\end{figure}

Figure 5(a) shows that two contributions to \Chi, the narrow quasielastic  and high energy components, are in excellent agreement with the AIM/NCA calculations with \Vkf\ = 443 meV at $x = 0$ and 477 meV at $x = 0.5$. It appears that there is a nearly linear relation between \Vkf\ and $x$, although there is no \textit{a priori} reason for this to be so. The important result is that both the low energy and high energy regions of scattering have absolute cross sections that are accurately predicted at each value of \Vkf\ over the entire transition from heavy fermion to mixed valence states. 

On the other hand, Figure 5(b) shows that the measured intensities in the intermediate energy region, which, at lower values of $x$, includes the crystal field excitation, are much lower than predicted if we use the same values of  \Vkf. Reasonable agreement  between theory and experiment can only be achieved in this energy range if we increase the value of \Vkf\ by approximately 20 meV at each value of $x$. To explain this apparent discrepancy, it is important to know that, because the crystal field potential is diagonal in $J_z$,  the quasielastic response is purely longitudinal, whereas the crystal field excitation is purely transverse in character. The anomaly in the crystal field intensities therefore implies that the \textit{k-f} hybridization in \CeAl\ is anisotropic.

In earlier investigations, we used the Anisotropic Kondo Model (AKM) to explain the anomalous spin dynamics in Ce$_{1-x}$La$_{x}$Al$_3$  \cite{Goremychkin:2000p51,Goremychkin:2002p30}. However, the present results provide the first direct evidence of the anisotropy in \Vkf\ and its impact on the dynamic magnetic susceptibility. There is some evidence that anisotropic hybridization may also be important in other strongly correlated electron systems, such as CeCu$_{6-x}$Au$_x$\cite{Lohneysen:2007p6354}, where the magnetic fluctuations appear two-dimensional, and URu$_2$Si$_2$ \cite{Elgazzar:2009p19226}, where we have proposed that the AKM plays an important role in generating hidden order \cite{Goremychkin:2002p30}.

In conclusion, a quantitative comparison of inelastic neutron scattering data with NCA calculations of the Anderson Impurity Model have provided, for the first time, a consistent description of the evolution of the spin dynamics from heavy fermion to mixed valence behavior. Because of the non-analytic character of the Kondo coupling, a 10\% increase in the \textit{k-f} hybridization produces an increase in the energy scale of spin fluctuations of more than an order-of-magnitude. However, the most significant result is that the transition between the two regimes does not primarily result from a general broadening of the dynamic magnetic susceptibility but by a significant shift of spectral weight from from the narrow components that characterize well-localized $f$-electron states to a broad magnetic response that is present at all hybridization strengths, including heavy fermion compounds such as CeAl$_3$. 

This research was supported by the U.S. Department of Energy, Office of Basic Energy Sciences, Division of Materials Sciences and Engineering under Award \# DE-AC02-06CH11357. 


\end{document}
\end